\documentstyle[NATO,epsfig]{crckapb}
\begin{opening}
\title{SUBSTRATE-MEDIATED INTERACTION ON AG(111) SURFACES FROM FIRST
PRINCIPLES}
\author{KRISTEN A. FICHTHORN}
\institute{The Pennsylvania State University\\
University Park, PA 16802 USA}
\author{MATTHIAS SCHEFFLER}
\institute{Fritz-Haber-Institut der Max-Planck-Gesellschaft\\
D-14 195 Berlin-Dahlem, Germany }
\end{opening}
\runningtitle{SUBSTRATE-MEDIATED INTERACTION}
\begin{document}

\begin{abstract}
When two or more atoms bind to a solid surface, the substrate can mediate an
interaction between them.  In this paper, we use density-functional theory to
quantify the substrate-mediated pair interaction between two adatoms on a
compressively strained Ag(111) surface and on unstrained Ag(111).  On the
strained surface, the elastic interaction is significant over the short
range and
leads to a net attraction between two adatoms.  However, at the longest
distances
probed, the interaction is primarily electronic and repulsive. The
repulsion can
be as strong at 50 meV, and it forms a ring-like structure around an atom.  On
unstrained Ag(111), the interaction is primarily electronic in origin and it is
weak relative to the interaction found on the strained surface.  We
calculate the
energy barrier for an isolated atom to diffuse in each of these systems.
For the
strained surface, the magnitude of the diffusion barrier is comparable to
that of
the adsorbate interaction.  We discuss the implications of our findings for
growth
at surfaces.

\end{abstract}

\section {Introduction}

Adsorbate-adsorbate interactions at surfaces play a key role in controlling the
structure and properties of interfacial materials.   Such
interactions can be classified as through-space (direct) or through-substrate
(indirect), although a clear distinction is not possible~\cite{bradshaw}.
Quantification of the indirect, substrate-mediated interaction between two
adsorbates is the primary subject of this paper.

Substrate-mediated interactions can have an elastic and an electronic
component.
Elastic interactions arise when an adsorbed atom perturbs the positions of
surrounding surface atoms~\cite{lau1}.  Additionally, an adsorbate may
perturb the charge density near the surface and alter the binding energies of
neighboring adsorbates - giving rise to an electronic
interaction~\cite{bradshaw,einstein1,einstein2,lau2,persson}.  The electronic
interaction is oscillatory in nature and, in the limit that the separation $d$
between two adsorbates is much greater than the inverse Fermi wavevector ${\bf
k}_{F}^{-1}$, the interaction energy
$\Delta E$ is predicted~\cite{lau2} to have the form
\begin{equation}
\Delta E \sim \frac {cos(2{\bf k}_{F}d)}{d^{x}} \quad ,
\label{eq-de-asymp}
\end{equation}
where $x = 2$ if the substrate has a surface state and $x = 5$ in the
absence of
a surface state.  Oscillatory, charge-density waves associated with the
electronic interaction have been imaged in scanning-tunneling microscopy (STM)
studies of noble-metal surfaces~\cite{briner,brune1}, which posses a surface
state.  For example, on Ag(111), ${\bf k}_{F} = 0.08$ \AA~\cite{brune1} and
charge
density waves can be imaged at distances greater than 70 \AA~\cite{brune1}
in STM
studies.

Electronic, substrate-mediated interactions should also be manifested at
``intermediate'' distances between the short separations required for chemical
bonds and the long separations for which Eq.~\ref{eq-de-asymp} is applicable.
For example, Einstein and Schrieffer showed in tight-binding calculations that
the intermediate-range interaction is present and is, in fact, more significant
than the long-range interaction~\cite{einstein1}.  Recent, low-temperature STM
studies also indicate that this is the case~\cite{repp} for Cu adatoms on
Cu(111).  The intermediate range is accessible to modern, electronic-structure
calculations.  In this study, we use
 density-functional-theory (DFT) calculations to quantify the
substrate-mediated interaction between two adatoms on a (compressively)
strained
and on an unstrained Ag(111) surface.  We discuss the ramifications of these
interactions for adsorbate diffusion and growth at surfaces.

\section {Model and Methods}

Total energies and forces are calculated using the DFT code described by
Bockstedte {\it et al.}~\cite{bockstedte97}, employing fully separable,
norm-conserving pseudo-potentials of the Kleinman-Bylander
form~\cite{kleinman}.
These were generated using the code developed by Fuchs and
Scheffler~\cite{fuchs99}.  For our calculations with Ag, the $s$
pseudopotential component is taken as local and we use $p$ and $d$ projectors.
The $4 d$-electrons are taken as valence states and their wave functions are
expanded in plane waves, which truncate at a cut-off energy of $E_{cut}=50$ Ry
for our calculations.  We use the generalized gradient approximation (GGA) of
Perdew, Burke, and Ernzerhof (PBE)~\cite{perdew96} for the exchange-correlation
functional.

The supercell approach is used to describe the surface.  For most results
presented here, the surface is modeled as a repeating $(4 \times 2)$ slab
with 16
atoms per unit cell and a thickness of four layers.   The vacuum spacing is
five
interlayer distances, or
$\sim$ 12\AA.  Adsorbed atoms are placed on one side of the slab and field
effects due to the asymmetry are correctly treated in the code by a
compensating
dipole field.  For some of the adsorbate interactions probed, it is
expedient to
utilize smaller slabs.  Thus, we also use four-layer
$(2 \times 2)$, $(4 \times 1)$, and $(2 \times 1)$ slabs for some of the
calculations. For all of the main results presented here, the ${\bf k}$-space
integration is performed with the equivalent of 4 special ${\bf k}$ points
in the
full, surface Brillouin zone of the $(4 \times 2)$ slab.  To improve
${\bf k}$-space integration, the electronic states are occupied according to a
Fermi distribution with $k_{B}T$=0.1 eV and the total energies are extrapolated
to zero temperature.

To verify the accuracy of the pseudo-potential, we calculated the lattice
constant of bulk Ag in its ground state.  For these calculations, we use
$E_{cut}=50$ Ry and 182 ${\bf k}$ points in the irreducible portion of the
Brillouin zone.  We find a lattice constant of $a_{0} = 4.20$\AA, which is
within
1 \% of the values found by Yu and Scheffler~\cite{yu} and by
Ratsch {\it et al.}~\cite{ratsch97}, who both used DFT with the GGA
functional of Perdew {\it et al.}~\cite{perdew92}.  As an additional test
of the
accuracy, we calculated the interlayer relaxation $d_{12}$ of the first layer
of Ag(111).  For these calculations, the first layer atoms are relaxed until
the forces on them are less than 0.005 eV/\AA.  We find that the first
layer relaxes inward by 2.0 \% relative to the bulk, interlayer spacing
$d_{b}$,
in good agreement with DFT local-density approximation (LDA) results
by Narasimhan and Scheffler~\cite{narasimhan}, who found an inward
relaxation of
1.7\%, and by Xie, {\it et al.}~\cite{xie}, who found an inward relaxation
of 1.0
\%.  Medium-energy ion scattering experiments indicate that $d_{12}$
contracts by
~2.5 \% relative to $d_{b}$ at temperatures below 670 K~\cite{statiris}.  A
more
recent low-energy electron diffraction investigation indicates no
contraction of
$d_{12}$ occurs on this surface~\cite{soares}. In previous theoretical studies
of Ag(111)~\cite{ratsch97,narasimhan,xie}, it was found that differences
between $d_{23}$ and $d_{b}$ are negligible.  Thus, in the calculations
presented
here, $d_{23}$ is fixed to $d_{b}$.

We study adsorbate interactions on both Ag(111) and on a strained Ag(111)
surface.  For the strained Ag(111) surface, we employ a bulk lattice
constant of
4.01 \AA.  This value is taken from the results of DFT-GGA calculations of the
lattice constant of Pt by Ratsch, Seitsonen and Scheffler~\cite{ratsch97}.  We
choose this substrate to address experimental issues related to Ag epitaxy on a
Pt(111) substrate covered by a monolayer (ML) of
Ag~\cite{ratsch97,ratsch98,brune95,fichthorn00}.  To describe the bare,
strained
Ag(111) substrate, the top layer of atoms is relaxed and the rest are fixed
to the
strained, bulk positions.  This results in an outward relaxation of the top
layer
by 9.0 \% relative to $d_{b}$ for the strained substrate.  The outward
relaxation
allows for relief of the strain induced by the bulk lattice, which is
compressed
by 4.61\% relative to bulk Ag, as it is described by our DFT calculations with
the GGA.

The total interaction energy $\Delta E$ for a periodic slab
 containing $N$ adatoms, of which $M$ are at binding site $a$ and $(N-M)$
are at
binding site $b$, is given by
\begin{equation}
\Delta E = E_{S+N}^{a,b} - ME_{S+1}^{a} - (N-M)E_{S+1}^{b} + (N-1)E_{S} \quad .
\label{eq-de}
\end{equation}
 Here, $E_{S+N}^{a,b}$ is the total energy of a slab with $N$ adatoms,
$E_{S+1}^{a}$ and $E_{S+1}^{b}$ are the total energies of slabs containing one
adatom, and $E_{S}$ is the total energy of a bare slab. We obtain $E_{S}$ from
the calculations described above. In an attempt to isolate the electronic
component of $\Delta E$, we obtain $E_{S+N}^{a,b}$, $E_{S+1}^{a}$,
and $E_{S+1}^{b}$ as follows:  A single atom is placed in
a binding site (an fcc or hcp three-fold hollow site) on a relaxed (as
described
above) bare slab and its height is optimized with respect to the fixed
substrate.
>From this calculation, we obtain $E_{S+1}^{a}$ and $E_{S+1}^{b}$.  To calculate
$E_{S+N}^{a,b}$, adatoms are placed in binding sites on the relaxed
(and fixed) substrate with heights fixed to values from the single-adatom
calculations. We use this procedure to obtain most of
the results described below.  Additionally, to resolve the elastic contribution
to $\Delta E$, we obtain $E_{S+N}^{a,b}$, $E_{S+1}^{a}$, and $E_{S+1}^{b}$ by
simultaneously relaxing the adatoms and the first-layer substrate atoms for
a few
trial configurations.

In the DFT supercell approach, the total interaction energy $\Delta E$ is
comprised of interactions between different adatoms in the slab and
interactions
between adatoms in the slab and the periodic-image adatoms.  We can express
$\Delta E$ in terms of these interactions using the lattice-gas Hamiltonian
approach (see, {\em e.g.}\cite{stampfl99}), which yields

 \begin{equation}
 \Delta E = \frac{1}{2}\sum_{i,j}V^{(2)}(d_{i,j}) \sigma_{i}\sigma_{j}
 +\frac {1} {3} \sum_{i,j,k}V^{(3)}(d_{i,j}, d_{i,k}, d_{j,k})
\sigma_{i}\sigma_{j}\sigma_{k} +
 \ldots  \quad .
 \label{eq-lattice-gas}
 \end{equation}
 Here, the summations run over all sites $i$ in the slab and all
 sites $j$ and $k$ in the supercell (which includes both the slab and its
periodic images), $\sigma_{m}$ is unity if site $m$ ($m = i,j,k$) is
occupied and
zero, otherwise,  $d_{m,n}$ is the distance between sites $m$ and $n$,
$V^{(2)}(d_{i,j})$ is the pair interaction between two adatoms on sites $i$ and
$j$, and $V^{(3)}(d_{i,j}, d_{i,k}, d_{j,k})$ is the trio interaction between
three adatoms on sites $i$, $j$, and $k$.  We neglect higher-order
interactions. Another assumption implicit in Eq.~\ref{eq-lattice-gas} is
that the interaction between adatoms at a fixed distance is independent of
whether these atoms occupy fcc or hcp sites.  Actually, in the results
presented
below, interactions between two adatoms occupying the same type of binding site
are always obtained for adatoms on fcc sites.  We performed one test of this
assumption for two adatoms on hcp sites.  This calculation showed that the
interaction was identical to that found for atoms on fcc sites.  Finally, in
calculations designed to resolve the electronic interaction (which comprise the
bulk of the results presented here), the adatom binding energies on fcc and hcp
sites are virtually equal: The fcc site is favored by less than 3 meV.

\begin{figure}
\epsfysize=9.7cm \centerline{\epsfbox{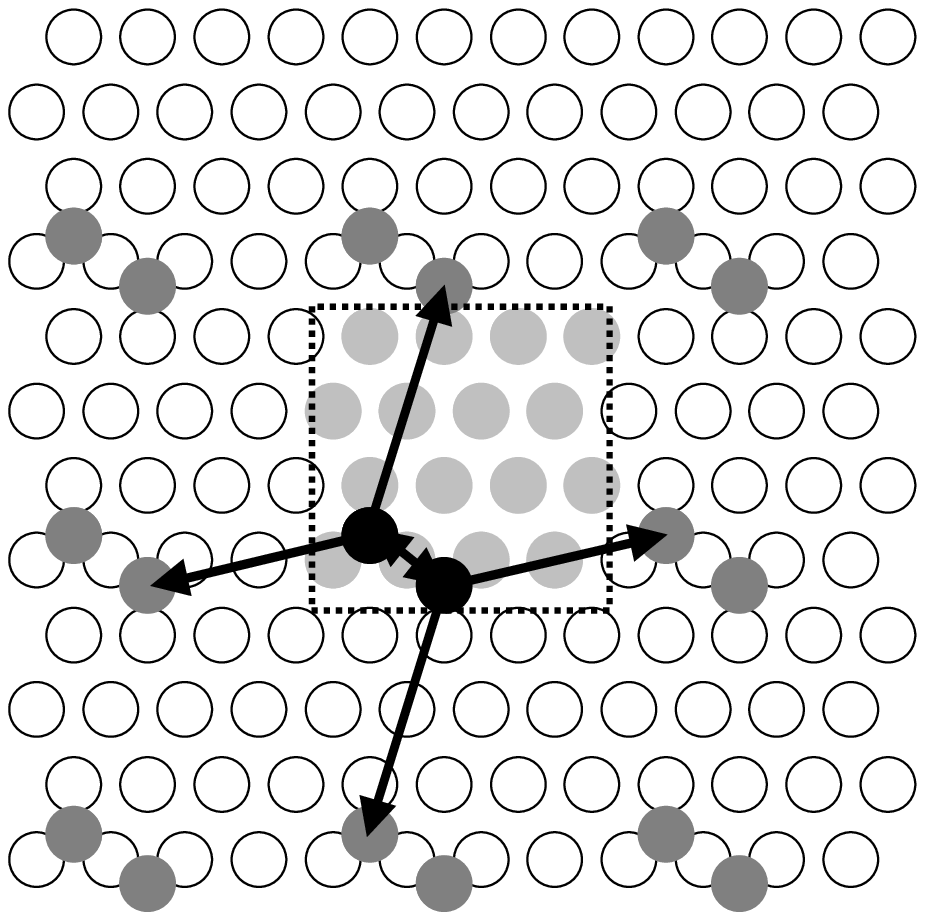}} \medskip
\caption{Sample lattice-gas Hamiltonian configuration.  Atoms in the slab are
enclosed in the dashed box.  The lines with arrows indicated relevant distances
between adatoms in the slab and periodic-image adatoms.}
\label{fig-lgh}
\end{figure}

To illustrate our approach, we show one of the structures probed in our
calculations in Fig.~\ref{fig-lgh}.  In this structure, two adatoms are in
the $(4 \times 2)$ surface unit cell at a separation of 3.43 \AA~ (distance
2, as
can be inferred from Fig.~\ref{fig-pair} below).  Each of these adatoms
has two, neighboring, periodic-image adatoms at a distance of 9.08 \AA~
(distance
12 shown on Fig.~\ref{fig-pair} below).  From Eq.~\ref{eq-lattice-gas},
$\Delta E$
for this configuration is comprised of one pair interaction at distance 2
and two
at distance 12.  Thus, for a given adatom configuration, we express $\Delta
E$ as
a sum of pair and trio interactions with unknown coefficients.   From 13
different
configurations, we obtain a system of linear equations and solve these for
pair-interaction coefficients  up to the $13^{\rm th}$-neighbor.  We assume
that
all other interaction coefficients are zero.

\section {Results and Discussion}

Before proceeding to our results, we discuss our tests of their accuracy.  We
test the convergence of our results with respect to $E_{cut}$, the number of
${\bf k}$ points, the slab thickness, and the exchange-correlation
functional for
different adsorbate configurations on Ag(111) and strained Ag(111).  The
adsorbate
configurations in these tests are chosen for two major reasons.  First, for
each
configuration, the total interaction energy is composed of interactions at a
single, adsorbate-pair distance.  Thus, it is possible to extract pair
interactions without solution of the full system of equations for the
lattice-gas
Hamiltonian.  A second reason for choosing these configurations is that they
posses a single mirror plane - the highest symmetry that could be achieved
here.  The (relatively) high symmetry makes it possible to retain up to 16
${\bf
k}$ points in the full surface Brillouin zone of a $(4 \times 2)$ unit cell, so
that we can assess the convergence of the ${\bf k}$-point summations for these
configurations.  For configurations of lower symmetry, 4 ${\bf k}$-points
in the
$(4 \times 2)$ unit cell is the maximum that we could use with the
computational
resources available to us.  In Table 1 we show the influence of the cut-off
energy, the slab thickness, and the number of ${\bf k}$ points on three,
pair-interaction coefficients for Ag adatoms on Ag(111).  The distances for the
pair interactions can be seen below in Fig.~\ref{fig-pair}.  For each result in
Table 1, each of the three energies necessary to obtain the pair-interaction
coefficient (cf., Eq.~\ref{eq-de}) is obtained with the indicated cut-off
energy,
slab thickness, and number of ${\bf k}$ points.
\begin{table}[htb]
\begin{center}
\caption{Convergence tests for $V^{(2)}(d)$.}
\begin{tabular}{ccccc}
\hline
Distance, $d$ & $N_{{\bf k}}$ & $E_{cut}$, Ry & Layers & $V^{(2)}(d)$, eV \\
\hline
    5 &  4 &     50 & 4 & -0.012 \\
    5 &  4 &     60 & 4 & -0.015 \\
    5 &  4 &     50 & 5 & -0.005 \\
    5 & 16 &     50 & 4 & 0.012 \\
    4 &  4 &     50 & 4 & 0.004 \\
    4 & 16 &     50 & 4 & 0.001 \\
    9 &  4 &     50 & 4 & 0.005 \\
    9 & 16 &     50 & 4 & -0.001 \\
\hline
\end{tabular}
\end{center}
\end{table}

From Table 1, we see that the tests for distance 5 indicate satisfactory
convergence with respect to the cut-off energy and the slab thickness.  The
results of our ${\bf k}$-point test for distance 5 are less satisfying and
there
is a (relatively) large difference between the pair-interaction coefficients
obtained with 4 and 16 ${\bf k}$ points.  To further probe the convergence
of our
results with respect to the ${\bf k}$-point summation, we investigated two
additional adsorbate configurations containing distances 4 and 9.  For these
configurations, differences in the pair-interaction coefficients are small and
indicate a satisfactory ${\bf k}$-point convergence.  It appears that
distance 5
is somehow unique.  We will discuss this configuration in more detail below.

To assess the dependence of our results on the exchange-correlation functional,
we obtain pair-interaction coefficients for three, different adatom
separations on strained Ag(111) using the LDA~\cite{ceperly}. The
pseduopotentials
for these calculations are generated as described above for the GGA
calculations,
except that the LDA is used for the exchange-correlation functional. We
test this
pseudopotential by calculating the bulk lattice constant for Ag using the same
conditions as for the GGA calculations.  We find $a_{0}$ = 4.06 \AA, which
is in
exact agreement with the value found by Narasimhan and
Scheffler~\cite{narasimhan} and within 1\% of the value found by
Ratsch, {\it et al.}~\cite{ratsch97}.  For strained Ag(111), we employ a
lattice
constant with a value of 3.92 \AA, as found by Ratsch {\it et
al.}~\cite{ratsch97}
in DFT-LDA calculations of the lattice constant for bulk Pt.

\begin{figure}
\epsfysize=8.8cm \centerline{\epsfbox{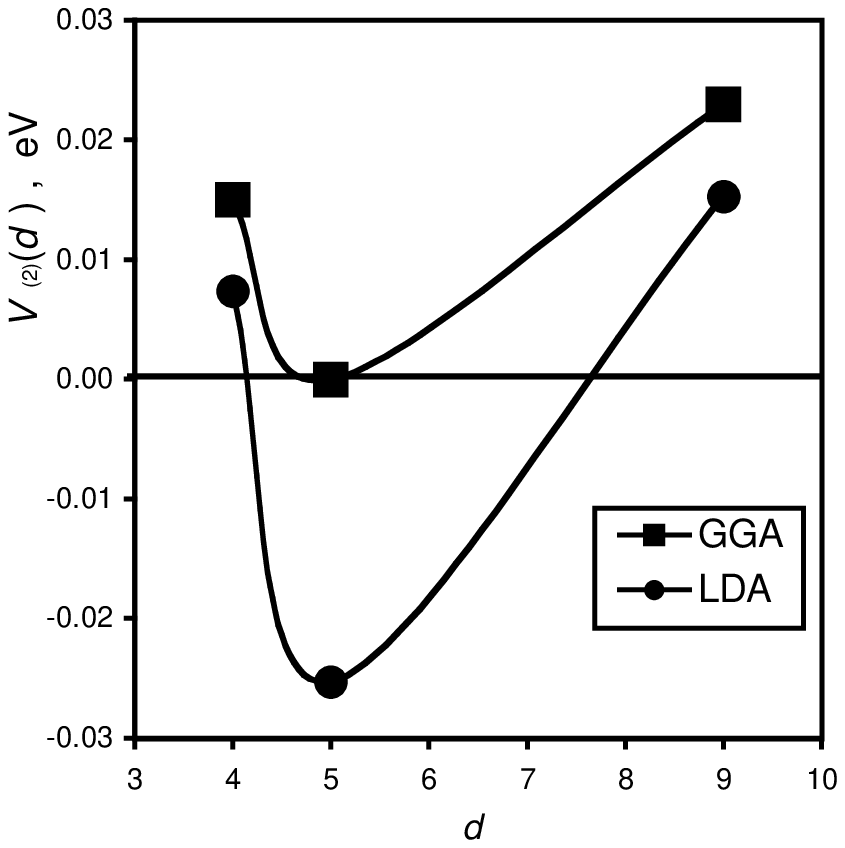}}
\caption{Comparison of LDA and GGA pair-interaction energies $V^{(2)}(d)$ for
three different adsorbate separations $d$.}
\label{fig-lda-gga}
\end{figure}

In Fig.~\ref{fig-lda-gga}, we show a comparison of the LDA and GGA results.
Here,
we see a small difference between the pair-interaction coefficients for
distances
4 and 9 and a (relatively) large difference between the LDA and the GGA at
distance 5.  Some of the discrepancy between the LDA and GGA results can
likely be
attributed to differences in strain.  The lattice in the GGA calculations is
compressively strained by 4.61 \% relative to the bulk, while the strain
for the
LDA system is 3.21 \% (for a reference, the experimental strain is 4.2
\%~\cite{brune95}).  As we will discuss in more detail below, we find that pair
interactions on Ag(111) follow the same trend as those on compressively
strained
Ag(111), but they are consistently lower.  A comparison of the LDA and GGA
results on Fig.~\ref{fig-lda-gga} also indicates this trend.  Thus, it is
likely
that strain plays some role in the differences between the LDA and the GGA
results.

\begin{figure}
\bigskip \epsfysize=4.7cm \centerline{\epsfbox{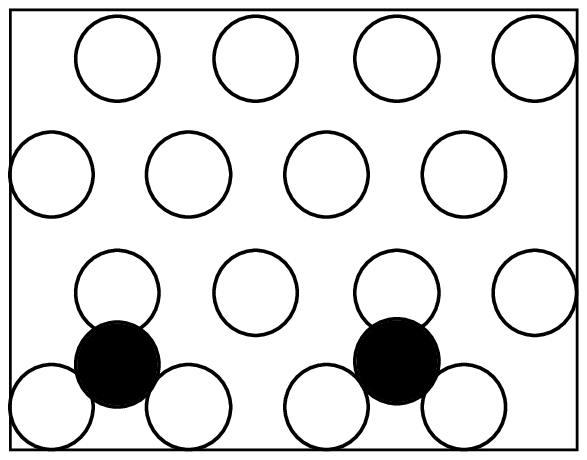}}
\caption{Configuration of adatoms used to obtain $V^{(2)}(5)$.}
\label{fig-config5}
\end{figure}

 We now return to our discussion of $V^{(2)}(5)$, for which we observe
large discrepancies with different ${\bf k}$-point summations (Table 1) and
with
the LDA vs. the GGA (Fig.~\ref{fig-lda-gga}).  Although the exact reason
for this
large error bar is unclear, one possible explanation is that trio interactions
contribute to the total interaction energy of the configuration that we used to
obtain $V^{(2)}(5)$.  As we see in Fig.~\ref{fig-config5}, this configuration
consists of a row of atoms occupying alternating fcc sites along the
$\langle 110
\rangle$ direction.  A trio in this configuration would most likely consist of
three adatoms in alternating fcc sites along the $\langle 110 \rangle$
direction.  Although we are not able to quantify this particular trio
interaction, we know that the trio interaction $V^{(3)}(4,5,9)$ is
attractive by
12 meV~\cite{fichthorn2b}.  Einstein found in simple, tight-binding
calculations
that the magnitude of the trio interaction is determined primarily by the two,
shortest adatom separations in the trio~\cite{einstein3}.  In light of these
results, a non-zero trio interaction seems possible for the adatom
configuration
in Fig.~\ref{fig-config5}.  If this is the case, then one trio interaction is
incorporated into our pair parameterization for $V^{(2)}(5)$ and the combined
uncertainties in both interactions could lead to the relatively large error bar
for this interaction coefficient.

\begin{figure}
\epsfysize=18.0cm \centerline{\epsfbox{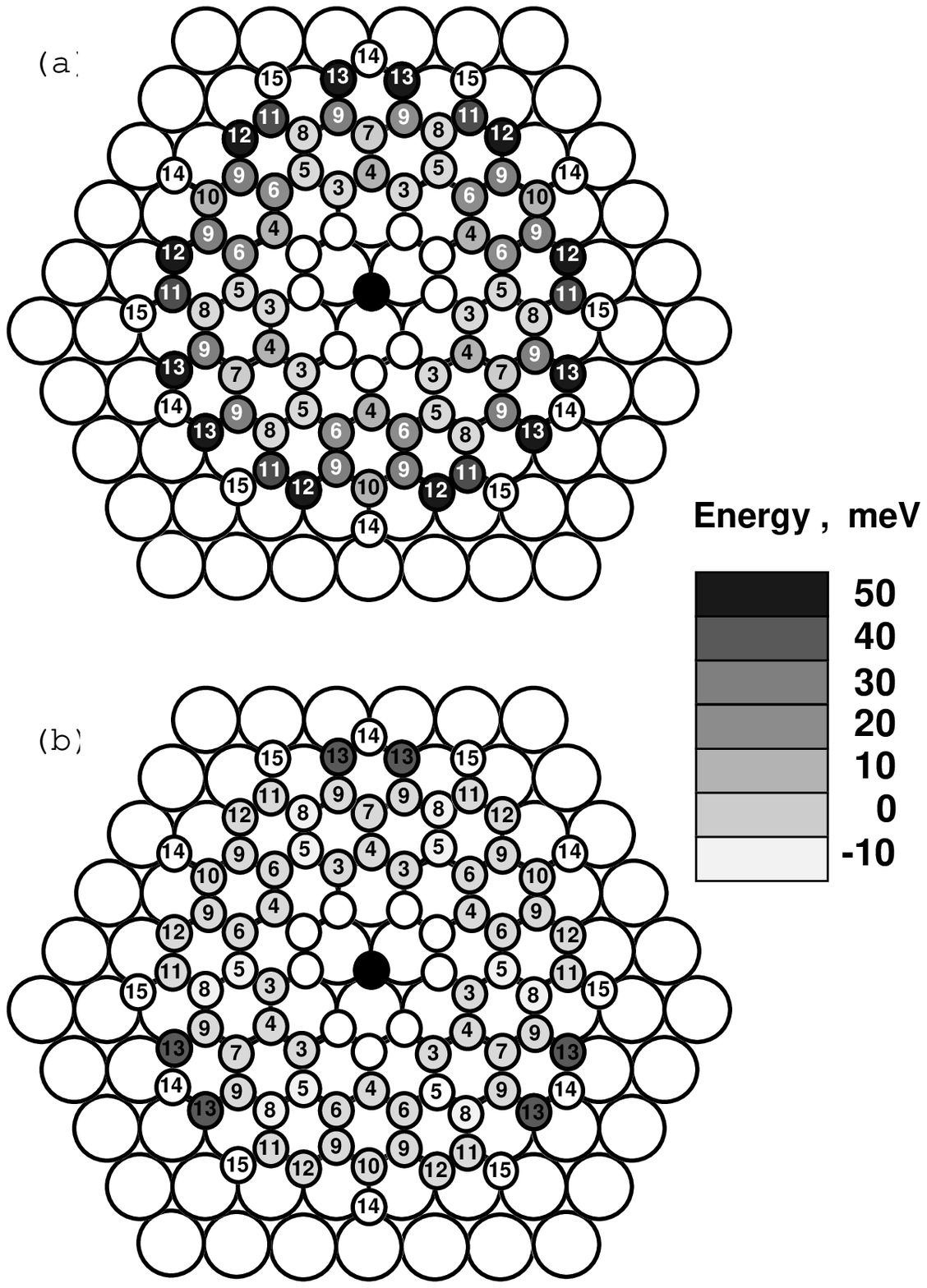}} \medskip
\caption{Pair interaction as a function of distance from a central adatom
(shown
in black) for strained Ag(111) (a) and unstrained Ag(111) (b).}
\label{fig-pair}
\end{figure}

Figure~\ref{fig-pair} shows the  pair interaction for the various distances
probed.  To facilitate resolution of the long-range interactions, the
short-range
interactions at distances 1 (the nearest-neighbor fcc site to the central
adatom)
and 2 ( the second-closest hcp site to the central adatom) are not shown.  For
both surfaces, the interaction is relatively weak at moderate distances and
it is
the strongest and the most repulsive at the longest distances probed.  For
strained Ag(111) [cf., Fig.~\ref{fig-pair}(a)], the interaction can be as
repulsive as $\sim 50$ meV at the longest distances and the repulsion forms a
ring-like structure around the central adatom.  The interactions are weaker on
Ag(111) [Fig.~\ref{fig-pair}(b)] and do not form a complete ring over the
distances probed here.  Generally, we find that the adatom pair interaction is
lower for Ag(111) than it is for strained Ag(111) - at all distances for which
the pair interaction is appreciable, the attractions on Ag(111) are
stronger and
the repulsion is weaker. We note that Bogicevic and co-workers~\cite{bogicevic}
have recently obtained interactions similar to ours in DFT-LDA calculations
of Al
on Al(111) and Cu on Cu(111).

Since the surface atoms have the same configuration in all of the results
discussed above, the elastic interaction due to adsorbate-induced rearrangement
of the surface atoms is suppressed.  To assess the role of elastic
interactions,
we recompute the pair-interaction coefficients in a few trial structures for
which both the adsorbate and first-layer surface atoms are fully relaxed until
the forces are typically less than 0.025 eV$/\AA$.

In Table 2, we show the elastic contribution to the pair-interaction
coefficients
at several adsorbate separations.  This is given as $\Delta V^{(2)}(d)$, where
$\Delta V^{(2)}(d)$ is the difference between the pair-interaction coefficient
obtained for a fully relaxed structure and that for the electronic interaction.
>From Table 2, we see that the elastic interaction plays a very small role
in the
total interaction energy for Ag(111).  The largest relaxation is for a dimer, a
weak attraction of 14 meV.  The elastic interaction is considerably
stronger for
strained Ag(111) and is seen to be appreciable up to distance 5 on
Fig.~\ref{fig-pair}.  At distances 4 and 5, the reduction of the pair
interaction upon relaxation is greater than the electronic interaction [cf.,
Fig.~\ref{fig-pair}(a)], indicating that the elastic interaction is dominant at
these distances.  At longer distances, the elastic interaction is weak and we
conclude that the repulsive ring seen on strained Ag(111) is electronic in
origin.

\begin{table}[htb]
\begin{center}
\caption{The elastic contribution $\Delta V^{(2)}(d)$ (units of eV) to the pair
interaction between two adsorbates separated by a distance $d$. }
\begin{tabular}{ccc}
\hline
 Distance, $d$ & $\Delta V^{(2)}(d) -$ Ag(111) & $\Delta V^{(2)}(d) -$ Strained
Ag(111) \\
\hline
1  & -0.014 & -0.051  \\
4  &  0.003 & -0.016  \\
5  &  0.003 & -0.023  \\
9  &  0.003 & -0.007  \\
11 &  0.003 & -0.003  \\
\hline
\end{tabular}
\end{center}
\end{table}

Adsorbate interactions at surfaces can influence both thermodynamically favored
structures and adsorbate diffusion.  To assess the impact of adsorbate
interactions on surface diffusion, we obtain the diffusion barrier for an
isolated adatom to hop between adjacent three-fold hollow sites on both
surfaces.  In these calculations, both the adatom and the surface atoms are
relaxed.  The diffusion barrier is given by the difference between the total
energy when the adatom is at a bridge site and when the atom is at a fcc
three-fold hollow site. We find for Ag(111) that the diffusion barrier is 87
meV.  For strained Ag(111), we find a barrier of 52 meV.   Our barriers are in
good agreement with those of Ratsch {\it et al.}~\cite{ratsch97}, who found 81
meV for Ag(111) and 63 meV for strained Ag(111) in DFT-LDA calculations.
Differences between the barriers that we find and those found by Ratsch can be
attributed to differences in the exchange-correlation functional (LDA vs.
the GGA
used here).  Additionally, there are differences between the lattice constant
predicted for the strained Ag(111) surface in the two studies.  As discussed
above, the lattice in our GGA calculations is compressively strained by 4.61 \%
relative to the bulk, while the strain for the LDA system is 3.21 \%.  Ratsch
{\it et al.} found that the value of the diffusion barrier on a strained
Ag(111)
surface is very sensitive to the degree to which the surface is strained
and that
smaller diffusion barriers arise on surfaces under greater
compression~\cite{ratsch97}.  The difference between our diffusion barrier and
that found by Ratsch {\it et al.}~\cite{ratsch97} is adherent to this
trend.  Our
diffusion barriers are also in good agreement with experimental
values~\cite{brune95} for Ag on 1-ML-Ag/Pt(111) (60 meV) and on Ag(111) (97
meV).

Comparing the diffusion barrier for an adatom on strained Ag(111) to the
magnitude of the pair interactions on this surface, we see that the two are
comparable.  Thus, one conclusion arising from this work is that
substrate-mediated interactions can influence adsorbate diffusion
and morphologies formed during growth at surfaces.  Our recent kinetic
Monte Carlo
simulation studies of island nucleation and growth show that this is the
case~\cite{fichthorn00,fichthorn2b}.  Recent experimental STM studies have
begun
to probe the role of adsorbate lateral interactions in determining the
structure
of adsorbate islands at surfaces~\cite{repp,besenbacher}.
Theoretical-experimental interplay in this area would be fruitful.

\section {Acknowledgements}

This research is supported by the Alexander von Humboldt Foundation and NSF
grant
No. DMR-9617122.


 \end{document}